\documentstyle[emulateapj,psfig]{article}

\newcommand{\be}{\begin{equation}}
\newcommand{\ee}{\end{equation}}
\newcommand{\ba}{\begin{eqnarray}}
\newcommand{\ea}{\end{eqnarray}}
\newcommand{\siml}{\lower4pt \hbox{$\buildrel < \over \sim$}}
\newcommand{\simg}{\lower4pt \hbox{$\buildrel > \over \sim$}}

\slugcomment{Accepted for publication in ApJ Letters}

\begin{document}

\title{Electromagnetic signals from planetary collisions}

\author{Bing Zhang \& Steinn Sigurdsson}
\affil{
Department of Astronomy \& Astrophysics, Pennsylvania State
University, University Park, PA 16802}

\begin{abstract}
Planet--planet collisions are expected during the early stages of the 
formation of extra-solar planets, and are also possible in mature
planetary systems through secular planet-planet perturbations.
We investigate the electromagnetic signals accompanied with these
planetary collisions and their event rate, and explore the possibility
of directly detecting such events. A typical Earth--Jupiter collision
would give rise to a prompt EUV-soft-X-ray flash lasting for hours and
a bright IR afterglow lasting for thousands of years. It is suggested
that with the current and forthcoming observational technology and
facilities, some of these collisional flashes or the post-collision
remnants could be discovered. 
\end{abstract}

\keywords{Planetary systems}

\section{Introduction}

More than 100 extra-solar planets have been detected. 
The growing sample allows us to study the properties of these
extra-solar planetary systems, and sheds light on the origin of our
own solar system. At the same time, our understanding of astrophysical
phenomena has been greatly boosted through studying cataclysmic
transient events such as supernovae, X-ray bursts, gamma-ray bursts
and their afterglows. Here we discuss another type of electromagnetic
transient events that arises from collisions of extra-solar
planets. Such giant impacts have been long predicted within the
theories of planet formation (e.g. Wetherill 1990). 
Giant impacts have been invoked to interpret the formation of
the Earth-Moon (Wetherill 1985; Benz, Slattery \& Cameron 1986; Canup
\& Asphaug 2001) and the Pluto-Charon (Stern 1992 and references
therein) systems, as well as the anomalous obliquities of Uranus and
some other solar giant planets (e.g. Slattery, Benz \& Cameron
1992). Collisions are also expected in mature planetary systems due to
slow onset, secular dynamical instabilities, 
although these collisions are less frequent than in dynamically young
systems. The probability of collisions is enhanced in white dwarf
planetary systems because of the perturbation caused by drastic
mass loss during the formation of the white dwarfs (Debes \&
Sigurdsson 2002). Accompanying an impact event, a prompt transient
electromagnetic signal and a fading afterglow are expected, but 
these signals have so far been sparsely studied. Even before
the discovery of planetary systems around main sequence stars,
Stern (1994) discussed the frequency and infrared (IR) signals of
planetary collisions.  In this Letter, we will sketch the likely
electromagnetic signals from planetary collisions (\S2). We will
estimate the possibility of directly detecting such collisional events
and their remnants, and discuss the observational strategies (\S3).

\section{EUV-soft-X-ray flashes and IR afterglows}

As an example, we consider a Jovian planet (primary) with mass $M_1 =
10^{30}~{\rm g}~ m_J$ and radius $R_1=7.0\times 10^9~{\rm cm}~ r_J$
being hit by a terrestrial planet (secondary) with mass $M_2=6\times
10^{27}~{\rm g}~ m_\oplus$ and radius $R_2=6.4\times 10^8~{\rm cm}~
r_\oplus$, where $m_J$ and $r_J$ are the mass and radius in units of
the Jupiter values, while $m_\oplus$ and $r_\oplus$ are the mass and
radius in units of the Earth values, respectively. The reasons to
choose a Jupiter-sized object being the primary are several folds. 
(1) Over 100 extrasolar giant planets (EGPs) have been detected, so
that we know they are definitely there; (2) The signal is the
strongest for these impacts because of their large masses (and hence
higher Eddington luminosities); (3) Because of their larger
gravitational potential and larger cross-sectional area, the chance
of collision per planet is higher at higher masses. 
Since the main emission quantities scale with masses and radii, the
discussion below can be straightforwardly generalized to collisions
between planets with other parameters (e.g. EGPs). 

The probability of a glacing event is $\sim R_2/R_1$ for a
gravitational-focused system, or $\sim (R_2/R_1)^2$ when the colliding 
velocity largely exceeds the escaping velovity of the primary. For our
parameters, the case is marginal for both regimes, and the probability
for glacing is between $\sim 9\%(r_\oplus/r_J)$ and $\sim
1\%(r_\oplus/r_J)^2$, which is negligible. Hereafter we will
assume head-on collisions throughout\footnote{We note that for
collisions between planets with similar masses, the chance of glacing
events is much larger, which leads to warm planets rather than the
signals discussed in this Letter.}. Assuming zero velocity at
infinity, the total energy of the collision is 
$E \sim G M_2 M_1/R_1 \sim 6\times 10^{40} ~ {\rm erg}~ (m_\oplus
m_J/r_J)$, and the relative velocity upon impact is $v \sim 45~{\rm
km~s^{-1}}~(m_J/r_J)^{1/2}$. 
Earthquake data reveal that the average p-wave speed interior of the 
earth is $\sim 10~{\rm km~ s^{-1}}$ (Dziewonski \& Anderson 1981), so
we can generally define a sound speed $C_s \sim 
10~{\rm km~ s^{-1}} c_{s}$, with $c_{s} \sim 1$ for
terrestrial planets. Since $v \gg C_s$, strong shocks will 
propagate into both colliders, converting most of the kinetic
energy into internal energy and accelerating non-thermal electrons.
The shock crossing time for the impactor, i.e., the 
terrestrial planet in our case, is
\be
\tau_{sh} \sim  R_2/C_{s} \sim 640 ~{\rm s}~
(r_\oplus/c_{s}).  
\ee 
This defines the rising timescale of the collision-induced
electromagnetic flash. The energy deposit rate during this stage is
$\dot E \sim E/\tau_{sh} \sim 9\times 10^{37}~{\rm erg~s^{-1}}
(m_\oplus m_J/r_J) (r_\oplus /c_{s})$. This is 
$\sim 1.8\times 10^3$ times the Eddington luminosity
of the Jovian planet, $L_{\rm Edd} = 4\pi cG m_pM/\sigma_T \sim
5\times 10^{34} {\rm
erg~s^{-1}} m_J$ (where $c$, $G$, $m_p$, $\sigma_T$ are the speed of
light, gravitational constant, proton mass, and Thomson cross section,
respectively).  About one half of the collisional energy is
converted to internal energy, of which only a small fraction is
radiated promptly, while a large fraction is sunk as latent heat.
The other half of the energy is initially in kinetic form, driving a
radiation-supported expanding envelope and inducing oscillation and
convection within the giant planet interior. This kinetic energy will
be thermalized later and radiated away over a longer period of
time along with the latent heat, as an IR afterglow (see
below). In the prompt phase, photons must be trapped by the high
opacity in the regions where the heat is initially deposited, and the
peak bolometric luminosity is
\be 
L_{pk} = \eta L_{\rm Edd} \sim 5\times 10^{34}~{\rm erg~s^{-1}}~ \eta 
m_J \sim 13 L_\odot \eta m_J, 
\label{Lpk}
\ee 
where $L_\odot = 3.8\times 10^{33}~{\rm erg~s^{-1}}$ is the solar
luminosity, $\eta \leq 1$ is a factor correcting for less than optimal
emission geometry and other radiative inefficiencies.  Since $\dot E
\gg L_{\rm Edd}$, $\eta \sim 1$ is likely achieved. The spectrum is
essentially thermal from a hot spot\footnote{The hot spot will
eventually expand via a subsonic excavation flow (Melosh 1989) or a
convective flow, but
the timescale for this to happen is $\sim 1$ day, much longer than
the flash timescale discussed here (eq.[\ref{tau1}]).} with radius
$\sim R_2$, with a non-thermal hard tail caused by 
Comptonization of the thermal photons by the non-thermal electrons
accelerated from the shocks. The typical thermal temperature
at the peak time is
\be
T_{pk}=\left(\frac{L_{pk}}{4\pi R_2^2 \sigma}\right)^{1/4} 
\sim 1.1\times 10^5 ~{\rm K}~ \eta^{1/4} m_J^{1/4} r_\oplus^{-1/2},
\ee
which is insensitive to both the planet mass and the unknown 
radiation efficiency (i.e. $\propto \eta^{1/4} m_J^{1/4}$). 
The peak frequency is $\nu_{pk}
= 6.6\times 10^{15} ~{\rm Hz}~ \eta^{1/4} m_J^{1/4}
r_\oplus^{-1/2}$ (wavelength $\sim 450 {\rm \AA}$ and energy $\sim 28$ 
eV). This is in the extreme ultraviolet (EUV) band. The peak flux is 
\be
F_\nu (pk) = \frac{h\sigma T^3}{2.8k} \cdot
\left(\frac{R_2}{D}\right)^2 
=60~{\rm \mu Jy}~ \eta^{3/4} m_J^{3/4} r_\oplus^{1/2} \left(\frac{D} {\rm 
10 kpc}\right)^{-2},
\label{fnupk}
\ee
where $h$, $k$, and $\sigma$ are the Planck's, Boltzmann's and
Stefan-Boltzmann constants, respectively. The hard non-thermal tail
is expected to extend into the soft-X-ray band.

For the thermal spectrum, the flux around $\nu$ could
be estimated as $F(\nu) \sim (R/D)^2 B_\nu(T) \nu$, where $B_\nu(T)$
is the Plank function, $R$ is the emission radius, and $D$ is the
distance to Earth. One can define a ratio $f(\nu) \equiv
{F(\nu,p)}/{F(\nu,*)}$ 
as the flux contrast between the planet and the star at a certain
frequency $\nu$, where $F(\nu,p)$ and $F(\nu,*)$
are the fluxes of the planet and the star in the band around $\nu$,
respectively. At the flash peak time, the EUV flare can greatly
out-shine a main-sequence host star later than B5, or a white dwarf
host star.
In the optical and IR bands, high contrasts are also achievable for
late-type host stars. For example, for a G2 (solar-type) star, the 
U, V, and I bands contrasts at the peak time are 
$f({\rm U, pk})\sim 0.2$, $f({\rm V,pk}) \sim 0.02$, and $f({\rm
I,pk}) \sim 0.008$, respectively. Even higher contrasts are achievable
when a cooler star is adopted. For $f\geq 0.01$, the flare 
is detectable with photometric monitoring of ground-based
telescopes (e.g. Borucki et al. 2001). 

Fixing a particular band, a lightcurve could be depicted. The rising
lightcurve is very steep, with a time scale of $\sim \tau_{sh}
\sim 10$ minutes. The decay time scale after the peak is
likely not very long, based on existing observational data of
other related phenomena, e.g. Comet Shoemaker-Levy
9 vs. Jupiter collision (Orton et al. 1995), X-ray bursts
in accreting neutron stars (Lewin, van Paradijs \& Taam 1993) and 
soft gamma-ray giant flares from magnetars (Feroci et al. 2001).
The prompt radiation pressure would drive a fraction of the 
impacting materials outwards, so that initially the Eddington-limited
photosphere expands and then cools. As the radiation pressure gradually 
drops, the photosphere contracts and the materials falls back
onto the Jovian planet surface. 
This fallback refreshes the surface heating, leading to
continually refreshed electromagnetic flashes with descending
magnitudes. A best guess is that the prompt flash will follow a
decaying tail 
extending up to several 10's of the rising time scale. The typical
timescale for the prompt flash during which the emission luminosity is
above, say $50\%$ of $L_{pk}$ could be estimated
\be
\tau_1 \sim 10 \tau_{sh} \sim 2 ~{\rm hr} ~ (r_\oplus /c_s),
\label{tau1}
\ee
with the averaged bolometric luminosity $\bar L_{bol}^{flash} \sim 10 
L_\odot \eta m_J$. This is the first important time scale in the
problem. 

The total radiated energy during this prompt flash stage
is $E^{flash} \sim \bar L_{bol}^{flash} \tau_1 \sim 3 \times
10^{38}~{\rm erg}$, which is much less than the total impact
energy, i.e. $\ll E \sim 6\times 10^{40}$ ergs. 
So the bulk of the
collision-deposited energy is {\em not} radiated promptly.
Rather, it is mixed into deep inside of the planet and
stored as heat. It is released during a much longer time scale through
a decaying thermal-radiation afterglow. Model simulations reveal
that at later times a Jovian planet follows an empirical bolometric
luminosity law (Black 1980; Saumon et al. 1996), i.e., $L/L_\odot
\sim 10^{-5} m_J^{2.35} t_6^{-1.22}$, and $T_{\rm eff} \sim 860 ~{\rm
K}~ m_J^{0.51} t_6^{-0.28}$, where $t_6$ is the age in unit of Myr,
and the relations are valid for $t_6 > 1$. Notice that the luminosity
decaying index $\sim 1.22 > 1$, so that the total energy released 
during $t> 10^6$ yr is negligible compared with the
energy released during the early epoch. At earlier times, generally we
expect a luminosity decaying law $L=L_0 (t/t_0)^{-a}$ with $a<1$,
where $t_0$ and $L_0$ are
the time and luminosity when such a steady fading phase begins.
The duration of the bright afterglow (the second timescale in the 
problem) can be estimated as
$\tau_2=\tau_{ag}=[(1-a){E}/{L_0 t_0}]^{{1}/{(1-a)}} t_0$.
The values of $t_0$, $L_0$ and $a$, are unknown, and
$a$ may also change during the decay. In principle, these 
should be calculated from the giant planet models. 
As a crude estimate, we take $(t_0,
L_0) = (1~{\rm yr}, 3.6\times 10^{-4} L_\odot)$. The $L_0$ value
corresponds to an initial temperature of $T_0\sim$2500 K (e.g. Stern
1994 and references therein) emitting from the full giant 
planet surface with the normalized radius $r_J=1$. 
Given such an initial condition, $a < 1/4$ is required in order to
have $L(t_6=1) > 10^{-5} L_\odot$. The bright afterglow duration is
therefore 
\be
1.4\times 10^3 ~{\rm yr}~ (m_\oplus m_J) r_J^{-3}
\left(\frac{T_0}{2500~{\rm K}}\right)^{-4}  \siml \tau_2 \siml 10^4
~{\rm yr},
\ee 
where the lower and the higher limits correspond to $a=0$ and $a=1/4$, 
respectively, and for the lower limit, the explicit dependences on the 
model parameters have been introduced. The peak of this bright
afterglow is in the IR band, and it is more favorably discovered in longer
wavelengths. For example, for a G2 host star, the typical I- and
K-band planet-to-star flux contrasts are $f({\rm I,ag}) \sim 2.6\times
10^{-4}$ and $f({\rm K,ag}) \sim 1.7\times 10^{-3}$ (see also Stern
1994).

Between the prompt flash and the onset of the later shallow-decay
long-term afterglow, 
the bolometric luminosity has to drop by 5 orders
of magnitude with a decay index $a>1$. This would nicely match the
$(t_0, L_0)$ initial condition adopted in the above discussion. 
In the above estimate we have assumed that the prompt flash radiation
is released from the hot spot, while the long-term afterglow emission
is emitted from the full surface. 
It is inferred that the Jupiter interior must be convective (Hubbard
1968; Bishop \& DeMarcus 1970). Thus the collision-deposited energy
should be rapidly transported to the planet bulk during tens or
hundreds of the convective overturn time scale. The depth into which
the impactor velocity is reduced to 1/2 is $\sim 2.5 \times 10^9$ cm,
i.e., the depth where the impacted mass equal to the mass of the
impactor, which is about 1/3 of the Jovian planet radius\footnote{The
penetration depth at which most of energy is deposited is even deeper,
especially when the mantle of the terrestrial planet is stripped
(Canup \& Asphaug 2001), but not much larger than the value quoted
here.}. This length scale may 
be adopted as the convective mixing-length. The convective speed is
less clear due to the unknown interior viscosity. Nonetheless it
should be subsonic, and we take $v_c \sim 1 ~{\rm km~s^{-1}} v_{c,0}$
for the following discussions. This gives a convective overturn
timescale $\tau_{conv} \sim 0.3 v_{c,0}^{-1}$ days. The timescale for
the excavative flows (Melosh 1989), if any, is longer than this.
Taking a typical overturn number of $N_c = 100 
N_{c,2}$ to achieve a full-surface isothermal state, we estimate a
third timescale 
\be
\tau_3 \sim 1 ~{\rm month}~ N_{c,2} v_{c,0}^{-1}.
\ee
Since the typical spin period of giant planets is $P_p < 1$ day
$\ll \tau_3$, one expects a periodic luminosity modulation on the
EUV-Soft-X-ray lightcurve, caused by the expanded hot spot (about  
$3\%$ of the full surface) entering and leaving the line-of-sight,
over a time scale of $\tau_3$.
This may be also detectable in the U, B bands or even in the V band
through photometry. 

During the vigorous convective epoch, it is likely that the
convection-induced turbulence would amplify the magnetic field to
higher values than conventional (e.g. sub-Gauss for Jupiter). 
The fast relativistic electrons accelerated
from the shocks or from magnetic reconnections would radiate
coherently, giving rise to a radio flare similar to the 
Type-II solar radio flares (but with much higher amplitudes). 
For a flare with brightness temperature $T_b > 10^{15}$ K, the
flux would out-shine the host star, and is $\sim 1\mu$Jy at a distance
of 10 kpc. Only very nearby events (e.g. within 1 kpc) are favorable
for radio observations. 

\section{Detectability and searching strategy}

Numerical simulations (Chambers, Wetherill \& Boss 1996; Ford,
Havlickova \& Rasio 2001) suggest that collisions of the type we are 
discussing are plausible. The basic picture is that there are secular
perturbations of the inner planets, which over time scales comparable
to the age of the system (Quinlan \& Tremaine 1992) lead to large
changes in eccentricity and semi-major axis for one or more planets,
leading to a large probability of collision. This is even favorable
for those EGPs that are found much closer to the host star than
Jupiter. Current theories conjecture that these EGPs migrate towards
the host star within the gas disk from which they were born (Lin,
Bodenheimer \& Richardson 1996). The disk dissappears in about 10
million years. Recent meteorite isotope studies suggest that
terrestrial planets form in a mean time scale of 10 million years
(Yin et al. 2002; Kleine et al. 2002; Jacobsen 2003), so that they
could be available for the collisions we conjecture\footnote{In case
of a late formation epoch for terrestrial planets (Wetherill 1985), a
proto-terrestrial planet gives a smaller total impact energy and a
shorter IR afterglow, but the prompt, Eddington-limited, flash
luminosity essentially remains the same (eq.[\ref{Lpk}]). }.

Considering an ensemble of stars with an average age $\bar t$, in
order to detect one 
event with duration $\tau$ (e.g. $\tau_1$ for the prompt flash and
$\tau_2$ for the afterglow) after a continuous observation time of
$t_{obs}$, the critical number of stars in this ensemble that have to
be searched is 
\be
N_{*} = (f_p \bar N_c)^{-1} \frac{\bar t}{{\rm max}(\tau,t_{obs})}, 
\ee
where $f_p$ is the fraction of the stars in the ensemble that have
planets, and $\bar N_c$ is the average total number of collisions
during the lifetime of a typical star in the ensemble. Evidence
suggests that $f_p$ could  
be close to unity (e.g. Lineweaver \& Grether 2003; Bary, Weintraub \&
Kastner 2002), 
and we will assume this most optimistic case.
$\bar N_c$ could be roughly estimated as the number of planets in the
dynamical system (which is greater at earlier epochs since each
dynamical interaction destroys one planet) multiplied by the
probability of direct collisions (rather than dynamical ejections),
which is in the $40\%$ to $80\%$ range (Ford et al. 2001). The high 
eccentricity of some EGP orbits may be caused by dynamical ejection 
of another EGP. Although no collision is involved in such events,
the eccentric orbits they result more likely lead to strong secular
perturbations of the remaining planets on a longer time scale,
particularly so if we assume inclinations are increased as well. These 
actually increase the probability of future collisions with planets
that might otherwise be unperturbed. A good portion of collisions
should happen in dynamically young systems (which associate with
star forming regions), but there could be also some collisions
occuring in mature planetary systems. For an old planetary system like
our Solar system, a typical value of $\bar N_c \sim 5$ may be
reasonable\footnote{Although the lack of a major planet in the
asteroid belt and the relatively small mass of Mars may be attributed
to Jupiter's perturbations that inhibited the growth of a
terrestrial planet near Jupiter required for collision, we note that
the solar system might be atypically stable (as compared with other
extrasolar systems discovered) for anthropic reasons.}, while for
dynamically young systems, one may adopt $\bar N_c \sim 3$.
A large fraction of stars are in binaries or multiples. Many of these
are either detached or very compact, allowing for the presence of
planets orbiting one or both stars (Holman \& Wiegert 1999; Hatzes et
al 2003). Planets may be less common around stellar binaries, but
those systems may contribute disproportionately to the net collision
rate, as a stellar companion can drive long term secular instabilties 
among multiple planet systems, increasing the collision rate per system.

Besides $N_*$, one can also define a characteristic flux $f_{\nu,c}$
of the collisional events. Given a number density $n_*$ of the stars
with the average age $\bar t$, one can estimate a critical distance
one has to search in order to find one collisional event, i.e., $D_c
\sim (N_*/n_*)^{1/3}$. The typical flux can be then estimated with
$D_c$. Although $N_*$ is sensitive to the average age $\bar t$ of the
ensemble of stars being investigated, $D_c$ and $f_{\nu,c}$ are
essentially independent of the ensemble adopted, given a constant
birth rate of stars. 

For the prompt flashes, since there are considerably fewer point
sources in the EUV and soft X-ray sky and since the collisional
events out-shine the host stars, one does not need to monitor the
individual stars in order to identify a collisional event. The
characteristic flux $f_{\nu,c}$ is relevant to discuss the
detectability. The {\em Extreme Ultraviolet Explorer (EUVE)} performed
a one-year EUV all-sky survey (Bowyer et al. 1994) and subsequently
conducted the serendipitous observations with the Right Angle Program
(RAP) (McDonald et al. 1994; Christian et al. 1999; Christian 2002)
for 8 more years. The RAP field of view is $\sim$ 20 square
degrees. So the combined observation is equivalent to a one-year
monitoring with $9\times(20/720)=1/4$ full-sky coverage. The total
number of stars with $\bar t \sim 10^{10}$ yr, for which one
collisional event is expected to have occurred in the {\em EUVE} field
of view, is $\sim 8\times 10^9$. The typical distance of such an event
is $\sim 10-20$ kpc, so that the critical peak flux is $f_{\nu,c} \sim
(15-60)\mu$Jy peaking around 450${\rm \AA}$. This is the band 
for which the ISM opacity is the largest. At shorter wavelengths where 
opacity is much lower, e.g., 50-100${\rm \AA}$ ($\sim 0.1-0.2$ keV),
the flux does not drop rapidly due to the non-thermal tail. Even for a
flux two orders of magnitude lower than the peak flux (a very
conservative estimate), i.e., $(0.15-0.6)\mu$Jy, this corresponds to
$(170-680)~{\rm counts~ks^{-1}}$ at 100${\rm \AA}$ for {\em EUVE}
(Bowyer et al. 1994), which is well above the survey sensitivity
threshold (McDonald et al. 1994). So at least one flare event with
such a flux level should have been detected. Since during 
$\sim 10$ yrs, $\sim 10\times 10^{11}/(10^{10}/5) \sim 500$
collisional events should have occurred in the Milkyway Galaxy 
($\sim 50$ per year), about 10 more events with lower flux levels
might have also been caught by {\em EUVE}, although with less
significance. The predicted flux level is also well above the
sensitivity of soft X-ray detectors, such as {\em ROSAT}, {\em
Chandra}, and {\em XMM-Newton}, so that there might be such events
recorded in their archival data as well. A future dedicated wide field
detector sensitive to (50-200)${\rm \AA}$ would be able to detect 10's
of EUV-soft-X-ray flares per year. The great opacity in the star
forming regions would reduce the flux level and even the
detectable event rate in the softer bands, but some events should be
directly detected, especially for those collisions outside of the star 
forming regions.

Another approach to identify prompt flashes is through
photometrically monitoring a huge number of stars over a long period
of time. As discussed above, during the flare, a 2-20\%
percent increase of the fluxes are expected for the late type
stars in the V band and bluer, which could be easily detected. Such a
photometric search could be achieved by future missions such as {\em
GAIA} (e.g. Perryman et al. 2001). Close monitoring $\sim 10^9$
stars for several years should lead to detection of several collisional
events.

The event rate of the IR afterglows is much larger thanks to their
much longer durations. Due to small flux contrasts ($\sim 10^{-3}$ in
K-band), one has to monitor the stars individually. The most economic
way is to search in the star forming regions (Stern 1994). For
an ensemble of stars 
with $\bar t \sim 5\times 10^7$ yr, one IR afterglow remnant would be
found by searching for $N_{*}^{ag} \sim 1.7\times 10^3 f_p^{-1} ({\bar
N_e}/{3})^{-1} ({\bar t}/{5\times 10^7}) \tau_{2,4}^{-1}$
young stars, where $\tau_{2,4}=\tau_2/10^4$ yr. The signals could be
identified through unresolved spectral measurements or direct
interferometric imaging (Stern 1994). Existing groundbased IR
interferometers have 1 mas resolutions (J. Ge, personal
communication), which corresponds to 0.13 AU at a distance of 130 pc,
the distance of the nearest star forming region. This is well adequate
to resolve bright planets from their host stars.

\section{Summary and discussion}

We have outlined the electromagnetic signals that would accompany with 
an event of collision between a Jovian planet and a terrestrial planet, 
and discussed the strategy of possibly detecting these events. 
The collision itself results in a bright EUV-soft-X-ray flare
lasting for hours. 
{\em EUVE} and some soft X-ray detectors might have detected some of
these flares. They may be also identified through monitoring a
huge number of stars with future missions such as {\em GAIA}, or
through future synoptic all--sky surveys.  After
detecting an EUV-soft-X-ray flare, a search of its optical counterpart
through photometric monitoring (like catching afterglows in gamma-ray
burst study) is desirable. A planetary-rotation-period
modulated fading signal would be an important clue. Doppler radial
velocity measurements and IR monitoring may be
performed later for the collision candidate to verify the existence of
the planet(s). Each collisional event would leave a bright remnant
glowing in IR. The duration of the afterglow is thousands of 
years. Such afterglows could be directly searched in the nearby star
forming regions, and the planets could be directly imaged. The current
planet-searching has been focused on the planets on stable orbits
(Perryman 2000). Searching for planetary collisional events will open a
new window in the planetary science. It allows us to witness the
gigantic events that are believed to have happened in the early history
of our solar system, and to directly test the planet-moon-formation
theories (see also Stern 1994). 

\acknowledgements 
We are grateful to Alan Stern, Damian Christian, Jian Ge, David
Weintraub and Kai Cai for 
helpful comments and discussions, and to the referee for insightful
comments that lead to improvement of the manuscript.
B.Z. acknowledges NASA NAG5-9192 for support. 
S.S. is supported by NSF grant PHY-0203046, the Center for
Gravitational Wave Physics (which is supported by the
NSF under co-operative agreement PHY 01-14375), and the Penn State
Astrobiology Research Center.

\end{document}